\documentclass[aps,prd,twocolumn,showpacs,amsmath,amssymb]{revtex4-1}

\usepackage[dvips]{graphicx}

\begin{document}


\title{Possible direct measurement of the expansion rate of the universe}
\author{Shi Qi}
\email{qishi11@gmail.com}
\author{Tan Lu}
\email{t.lu@pmo.ac.cn}
\affiliation
{
  Purple Mountain Observatory,
  Chinese Academy of Sciences,
  Nanjing 210008,
  China
}
\affiliation
{
  Joint Center for Particle, Nuclear Physics and Cosmology,
  Nanjing University---Purple Mountain Observatory,
  Nanjing 210093,
  China
}
\affiliation
{
  Kavli Institute for Theoretical Physics China,
  Chinese Academy of Sciences,
  Beijing 100190,
  China
}
\affiliation
{
  Key Laboratory of Dark Matter and Space Astronomy,
  Chinese Academy of Sciences,
  Nanjing 210008,
  China
}


\begin{abstract}
  A new method is proposed for directly measuring the expansion rate
  of the universe through very precise measurement of the fluence of
  extremely stable sources.
  The method is based on the definition of the luminosity distance and
  its change along the time due to the cosmic expansion.
  It is argued that galaxies may be chosen as the targets of the
  observation to perform the measurement.
  We show that, by simultaneously increasing the observation time
  and physically adding the fluences from different galaxies,
  the requirement on the relative precision of the detector for an
  observation of $1$ second on a single galaxy can be relaxed to
  $10^{-5}$.
  Benefiting from the abundance of galaxies in the universe, the
  method may be quite promising.
\end{abstract}



\maketitle

\section{Introduction}

Measuring the expansion rate of the universe along the redshift has
been one of the most important scientific objectives in cosmology
since the discovery of the cosmic expansion.
It is usually pursued by measuring the distances at different
redshifts, which, in turn, could be done with the data of the standard
candles like
the type Ia supernovae
(SNe Ia)~\cite{Phillips:1993ng},
Gamma-ray bursts
(GRBs)~\cite{Dai:2004tq, Ghirlanda:2004fs}
etc or the standard rulers from cosmic microwave background
(CMB)~\cite{Wang:2007mza}
and baryon acoustic oscillations
(BAO)~\cite{Bassett:2009mm}.
In fact, it is the measurement of the luminosity distances of
SNe Ia~\cite{Riess:1998cb, Perlmutter:1998np}
that leads to the discovery of the accelerating expansion of our
universe, which is attributed to the mysterious dark energy.
The further study of the nature of the dark energy requires more
precise expansion history of the universe.
Currently, the expansion rate of the universe is still mainly measured
through the distance measurement.
The distances depend on the expansion rate through an
integration, so the extraction of the expansion rate from the
distances involves differentials, which significantly affects the
precision of the measurement of the expansion, not to mention that the
derivation of the dark energy equation of state (EOS) from the
expansion rate involves differentials once again.

On the other hand, despite of the difficulties, some proposals have
been presented for directly measuring the expansion rate of the
universe, for example, through the measurement of radial
BAO~\cite{Hu:2003ti}, the relative ages of passively evolving
galaxies~\cite{Jimenez:2001gg}, the temperature and polarization
anisotropies of the CMB~\cite{Zahn:2002rr}, or the redshift
drift~\cite{Sandage:1962, Loeb:1998bu}
(the socalled Sandage-Loeb test).
In this paper, we propose another method to directly measure the
expansion rate by precisely measuring the fluence of extremely
stable sources.

\section{Methodology and discussion}

\begin{figure}[htbp]
  \centering
  \includegraphics[width = 0.5 \textwidth]{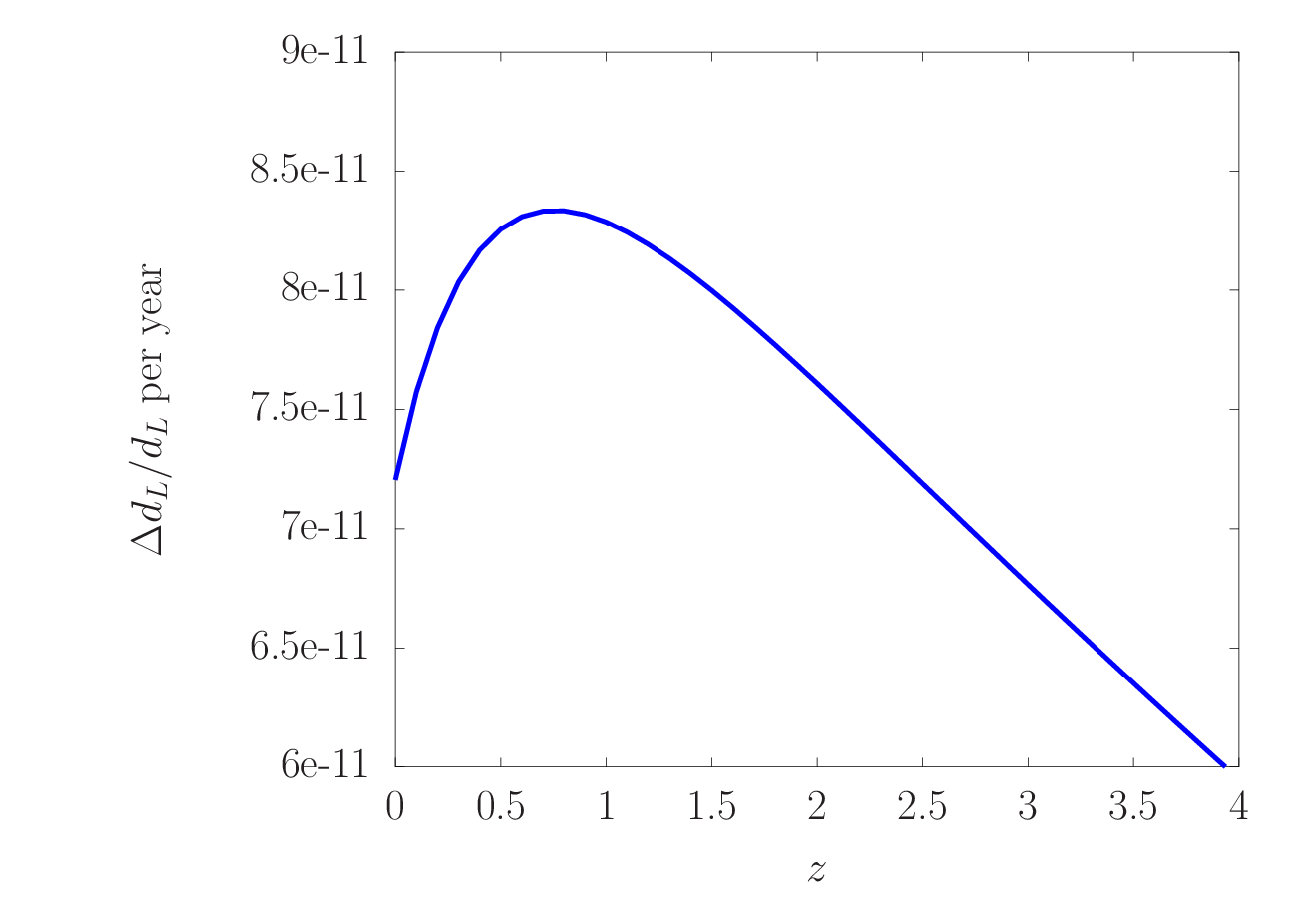}
  \caption
  {
    $\Delta d_L / d_L$ per year versus redshift $z$ for the flat
    $\Lambda$CDM cosmological model with $\Omega_{m, 0} = 0.27$ and
    $H_0 = 70.5 \ \mathrm{km} \, \mathrm{s}^{-1} \,
    \mathrm{Mpc}^{-1}$.
  }
  \label{fig:dlnd}
\end{figure}

Consider a source rest at comoving distance $r$, with a redshift of
$z$, its luminosity distance to us is given by
\begin{equation}
  \label{eq:luminosity_distance_1}
  d_L = a(t_0) r (1 + z)
  ,
\end{equation}
where $a$ is the scale factor as a function of time and $t_0$ denotes
the time of today.
We assume the signal we observed at the time of $t_0$ was emitted by
the source at $t_{\mathrm{em}}$.
Due to the expansion of the universe, if we observe the source again
after a time interval of $\Delta t_0$,
i.e.~at the time of $t_0 + \Delta t_0$
(the corresponding signal was emitted by the source at the time of
$t_{\mathrm{em}} + \Delta t_{\mathrm{em}}$
with $\Delta t_{\mathrm{em}} = \Delta t_0 / (1+z)$),
we will find its luminosity distance changed with a value (we only
take into account the first order terms in this paper) of
\begin{align}
  \label{eq:Delta_d_L}
  \Delta d_L
  & =
  \Delta a(t_0) r (1 + z)
  + a(t_0) r \Delta z
  \nonumber
  \\
  & =
  d_L \frac{\Delta a(t_0)}{a(t_0)}
  + d_L \frac{\Delta z}{1 + z}
  ,
\end{align}
where $\Delta a(t_0)$ and $\Delta z$ are the changes of $a(t_0)$ and
$z$ in the time interval $\Delta t_0$ due to the expansion of the
universe.
We can rewrite Eq.~(\ref{eq:Delta_d_L}) into
\begin{equation}
  \label{eq:Delta_lnd_1}
  \frac{\Delta d_L}{d_L}
  =
  \frac{\Delta a(t_0)}{a(t_0)}
  + \frac{\Delta z}{1 + z}
  .
\end{equation}
For the expansion of the universe, we have
\begin{align}
  \label{eq:Delta_lna}
  \frac
  {
    \Delta a(t_0)
  }
  {
    a(t_0)
  }
  & =
  \frac
  {
    \dot{a}(t_0)
  }
  {
    a(t_0)
  }
  \Delta t_0
  \nonumber
  \\
  & =
  H_0 \Delta t_0
\end{align}
and since $1 + z = a(t_0) / a(t_{\mathrm{em}})$,
\begin{align}
  \label{eq:Delta_z}
  \Delta z
  & =
  \frac
  {
    \dot{a}(t_0)
  }
  {
    a(t_{\mathrm{em}})
  }
  \Delta t_0
  -
  \frac
  {
    a(t_0)
  }
  {
    a(t_{\mathrm{em}})
  }
  \frac
  {
    \dot{a}(t_{\mathrm{em}})
  }
  {
    a(t_{\mathrm{em}})
  }
  \Delta t_{\mathrm{em}}
  \nonumber
  \\
  & =
  \frac
  {
    a(t_0)
  }
  {
    a(t_{\mathrm{em}})
  }
  \frac
  {
    \dot{a}(t_0)
  }
  {
    a(t_0)
  }
  \Delta t_0
  -
  (1 + z) H(z) \frac{\Delta t_0}{1+z}
  \nonumber
  \\
  & =
  \left[
    (1 + z) H_0 - H(z)
  \right]
  \Delta t_0
  ,
\end{align}
where $H_0$ and $H(z)$ are the Hubble parameter of today and that at
the redshift of $z$ respectively.
Eq.~(\ref{eq:Delta_z}) is in fact the core of the Sandage-Loeb
test~\cite{Sandage:1962, Loeb:1998bu}, i.e., if we manage to measure
the redshift drift $\Delta z$, we obtain the Hubble parameter.
Substituting Eq.~(\ref{eq:Delta_lna}) and Eq.~(\ref{eq:Delta_z}) into
Eq.~(\ref{eq:Delta_lnd_1}), we have
\begin{equation}
  \label{eq:Delta_lnd_2}
  \frac{\Delta d_L}{d_L}
  =
  \left[
    2 H_0 - \frac{H(z)}{1 + z}
  \right]
  \Delta t_0
  .
\end{equation}
With this equation at hand, one may naturally think that, similarly as
the Sandage-Loeb test, if the luminosity distances can be measured to
a very high precision such that we could distinguish the small changes
in the luminosity distances for a reasonable time duration, we could
also immediately derive the corresponding Hubble parameters.
Unfortunately, the distance measurement itself is a difficult task in
astronomy, especially for cosmic distances.
One of the most precise ways of measuring cosmic distances is
through the observation of gravitational waves, for which a relative
precision of about $10^{-3}$ is expected for the luminosity
distance~\cite{Arun:2008xf}.
From Fig.~\ref{fig:dlnd}, we can see that, even for such precise
measurements, about $10^8$ years will be needed before we could tell
the changes in the luminosity distances, which is obviously not
feasible.

Recall that the luminosity distance is defined through
\begin{equation}
  \label{eq:luminosity_distance_2}
  F = \frac{L}{4 \pi d_L^2}
  ,
\end{equation}
where $L$ is the luminosity of the source and $F$ is the
observed flux, we have
\begin{equation}
  \label{eq:Delta_lnF}
  \frac{\Delta F}{F}
  =
  \frac{\Delta L}{L}
  - 2 \frac{\Delta d_L}{d_L}
  ,
\end{equation}
where $\Delta F$ and $\Delta L$ are the changes in $F$ and $L$ during
the time interval $\Delta t_0$.
So, if we point our telescope to some extremely stable source such
that $\Delta L / L$ can be ignored in Eq.~(\ref{eq:Delta_lnF}) and
manage to measure the flux $F$ to a very high precision, we could
measure $\Delta d_L / d_L$ indirectly through measuring
$\Delta F / F$.
Thus, Eq.~(\ref{eq:Delta_lnd_2}) could still be used to measure the
expansion rate of the universe.
The problem becomes whether there exist such extremely stable sources
and whether we could manage to measure the small flux change.

For the requirement of extremely stable sources, we may consider
objects that include lots of similar sources, so that we could
statistically reduce $\Delta L / L$ to a very low level.
For example, a galaxy includes lots of stars.
During the time interval $\Delta t_0$, the luminosity of a star in the
galaxy may increase or decrease.
Let $L_i$ be the luminosity of the $i$th star in the galaxy, we may
view $\Delta L_i / L_i$ as a random variable and, for a simple
estimation, assume it follows the normal distribution
$\mathcal{N}(0, \, \sigma^2)$,
then, for the galaxy,
\begin{equation}
  \label{eq:Delta_lnL}
  \frac{\Delta L}{L}
  =
  \frac{\sum_i \Delta L_i}{\sum_i L_i}
  \sim
  \mathcal{N}(0, \, \frac{\sum_i L_i^2}{(\sum_i L_i)^2} \sigma^2)
\end{equation}
and since $\sum_i L_i^2 / (\sum_i L_i)^2$ has the order of $1/N$,
where $N$ is the total number of stars in the galaxy,
the standard deviation of $\Delta L / L$ for the galaxy is thus
reduced by a factor of about $1 / \sqrt{N}$ compared to that for a
star.
When considering the change in the luminosity, we first exclude
galaxies involving violent astrophysical processes like stellar
explosions, which usually could be easily identified.
If we simply model the luminosity evolution of a steadily burning star
to be of that it linearly increases to its maximum value and then
linearly decreases, then $\Delta L / L$ for the star should have the
order of the inverse of its lifetime.
The lifetime of a star ranges from only a few million years
(for the most massive) to trillions of years (for the least massive).
Here, we conservatively set the $\Delta L / L$ for stars to be of the
order of $10^{-6}$ per year which corresponds to stars with smallest
lifetime.
Typical galaxies consist of from $10^7$ to $10^{14}$ stars.
If we set $N = 10^{10}$, the standard deviation of $\Delta L / L$ for
the galaxy would be of the order of $10^{-11}$ per year.
So the $\Delta L / L$ can hopefully be reduced to below the expected
values of $\Delta d_L / d_L$.
In practice, some selection criteria on the galaxies may be needed.
The details on the selection are out of the scope of this short
paper. But it is worth mentioning that one may concern the impact
from the evolution of the observed galaxy beyond its stability.
On this issue, first we should, of course, select passively evolving
galaxies as our targets of observation basing our knowledge of
galaxies themselves.
Second, the $\Delta F / F$ caused by the cosmic expansion has a
flat-line spectrum, while those caused by other astrophysical
processes usually do not have such a character.
With this, we could further exclude those galaxies whose evolution
(including effects from star formation, dust, etc.)
dominates over the cosmic expansion on $\Delta F / F$.
In addition, the method itself does not impose any restriction on the
selection of the band, so we are free to choose the most appropriate
bands with least impacts from the evolution of galaxies to perform the
measurement.

For the measurement of the flux, cryogenic detectors usually can
achieve very high precisions.
See, for example~\cite{Peacock:1996, Day:2003, Enss:2005md}, for
cryogenic detectors.
However, despite these technologies, it is still a severe challenge
to the precision the instruments can achieve for directly measuring
the flux change caused by the cosmic expansion.
Here, with a simple trick, we further turn the measurement of the flux
to the measurement of the fluence and show that the requirement on the
precision of the instrument can be relaxed to an acceptable level.
Consider a sufficiently stable source, its flux changes due to the
expansion of the universe as
\begin{align}
  F(t)
  & =
  F(t_0)
  +
  \left.
    \frac
    {
      \mathrm{d} F
    }
    {
      \mathrm{d} t
    }
  \right\vert_{t = t_0}
  (t - t_0)
  \nonumber
  \\
  & =
  F(t_0)
  \left[
    1 +
    \left.
      \frac
      {
        \mathrm{d} \ln F
      }
      {
        \mathrm{d} t
      }
    \right\vert_{t = t_0}
    (t - t_0)
  \right]
  \nonumber
  \\
  & =
  F(t_0)
  \left[
    1 - 2
    \left.
      \frac
      {
        \mathrm{d} \ln d_L
      }
      {
        \mathrm{d} t
      }
    \right\vert_{t = t_0}
    (t - t_0)
  \right]
  \nonumber
  \\
  & =
  F(t_0)
  \left[
    1 - 2
    \left(
      2 H_0 - \frac{H(z)}{1 + z}
    \right)
    (t - t_0)
  \right]
  .
\end{align}
So the fluence as a function of time is
\begin{align}
  S(t)
  & =
  \int_{t_0}^t F(\tilde{t}) \mathrm{d} \tilde{t}
  \nonumber
  \\
  & =
  F(t_0)
  \left[
    (t - t_0)
    -
    \left(
      2 H_0 - \frac{H(z)}{1 + z}
    \right)
    (t - t_0)^2
  \right]
  .
\end{align}
For an observation time $T$, the fluence reaches $S(t_0 + T)$.
Basing on $T$ and $S(t_0 + T)$, we can construct a reference fluence
$S_T(t)$ that increases linearly to $S(t_0 + T)$, i.e.,
\begin{align}
  S_T(t) & = S(t_0 + T) \frac{t - t_0}{T}
  \nonumber
  \\
  & =
  F(t_0)
  \left[
    (t - t_0)
    -
    \left(
      2 H_0 - \frac{H(z)}{1 + z}
    \right)
    T (t - t_0)
  \right]
  .
\end{align}
Then we turn the problem to measuring the difference between $S(t)$
and $S_T(t)$, i.e.,
\begin{equation}
  S(t) - S_T(t)
  =
  F(t_0)
  \left(
    2 H_0 - \frac{H(z)}{1 + z}
  \right)
  \left[
    T (t - t_0) - (t - t_0)^2
  \right]
  .
\end{equation}
$S(t) - S_T(t)$ is a quadratic polynomial of time with its maximum
value at $t = t_0 + T/2$
\begin{align}
  \mathrm{Max}
  \left[
    S(t) - S_T(t)
  \right]
  & =
  S(t_0 + \frac{T}{2}) - S_T(t_0 + \frac{T}{2})
  \nonumber
  \\
  & =
  \frac{1}{4}
  F(t_0)
  \left(
    2 H_0 - \frac{H(z)}{1 + z}
  \right)
  T^2
  \nonumber
  \\
  & =
  \frac{1}{4}
  S(t_0 + T)
  \left.
    \frac{\Delta d_L}{d_L}
  \right\vert_{\Delta t_0 = T}
  .
\end{align}
Since $S_T(t_0 + T/2) = S(t_0 + T) / 2$, we have
\begin{equation}
  \label{eq:Delta_lnd_measure_1}
  \left.
    \frac{\Delta d_L}{d_L}
  \right\vert_{\Delta t_0 = T}
  =
  \frac
  {
    4 S(t_0 + \frac{T}{2}) - 2 S(t_0 + T)
  }
  {
    S(t_0 + T)
  }
  .
\end{equation}
Using $S_1$ and $S_2$ to denote the fluences during the first and last
half of the observation time, i.e.,
\begin{align}
  S_1 & = S(t_0 + T/2)
  ,
  \\
  S_2 & = S(t_0 + T) - S_1
  ,
\end{align}
in this case,
Eq.~(\ref{eq:Delta_lnd_measure_1}) can also be rewritten as
\begin{equation}
  \label{eq:Delta_lnd_measure_2}
  \left.
    \frac{\Delta d_L}{d_L}
  \right\vert_{\Delta t_0 = T}
  =
  \frac
  {
    2 (S_1 - S_2)
  }
  {
    S_1 + S_2
  }
  .
\end{equation}
So we only need measure $S_1$ and $S_2$ to derive
$\Delta d_L / d_L$.
Benefiting from this, we can simply add up fluences from
different sources at the same redshift, in addition to increasing the
observation time, to reduce the impact from the noise of the
instrument on the measurement.
Adding up the fluences also effectively increases the stability
of the source.
Denoting the fluences introduced by the noise of the instrument with
$\delta S$, the error of the measurement of $\Delta d_L / d_L$ caused
by the noise is
\begin{equation}
  \label{eq:measurement_error}
  \delta
  \left(
    \left.
      \frac{\Delta d_L}{d_L}
    \right\vert_{\Delta t_0 = T}
  \right)
  =
  \frac
  {
    2 (\delta S_1 - \delta S_2)
  }
  {
    S_1 + S_2
  }
  .
\end{equation}
Provided the noise level of the instrument does not change, the
standard deviation of $\delta S$ is proportional to the square root of
the observation time.
The relation between the standard deviation of $\delta S$ and the
number of independent sources added together, $N_s$, depends how the
addition is performed.
If the fluences are recorded independently for different sources and
added later by hand, the standard deviation of $\delta S$
would be proportional to $\sqrt{N_s}$.
Instead, if the fluences are physically added before the
photons reach the detector, such that the sources look as if it is one
source, but much brighter, to the detector
(for example, we can filter the light using a cover with a specific
pattern of small holes in it, so that only light from sources with a
specific redshift passes though, then we focus the light onto the
detector with a lens or a set of lenses),
the standard deviation of $\delta S$ would be the same as that for a
single source.
Obviously, the latter method, i.e., physically adding fluences from
different sources, is much more helpful to reduce the error
from the noise.
It is easy to check that the standard deviation of
Eq.~(\ref{eq:measurement_error}),
using this method, is $2 \epsilon (T / \tau)^{-1/2} N_s^{-1}$ under
the assumption for a simple estimation that the different sources have
the same brightness,
where $\epsilon$ is the relative precision of the detector
for an observation time $\tau$ on a single source.
So, to conduct any realistic measurement on $\Delta d_L / d_L$, it
requires
\begin{equation}
  \label{eq:requirement_1}
  2 \epsilon
  \left(
    \frac{T}{\tau}
  \right)^{-\frac{1}{2}}
  N_s^{-1}
  <
  \left.
    \frac{\Delta d_L}{d_L}
  \right\vert_{\Delta t_0 = T}
  .
\end{equation}
Since
\begin{equation}
  \left.
    \frac{\Delta d_L}{d_L}
  \right\vert_{\Delta t_0 = T}
  \sim
  6 \times 10^{-11}
  \frac{T}{1 \; \mathrm{year}}
  ,
\end{equation}
We have
\begin{equation}
  \label{eq:requirement_2}
  \epsilon
  <
  3 \times 10^{-11}
  \frac{T}{1 \; \mathrm{year}}
  \left(
    \frac{T}{\tau}
  \right)^{\frac{1}{2}}
  N_s
  .
\end{equation}
To get a general impression about how tight the requirement is,
setting $\tau = 1 \; \mathrm{second}$, we have
\begin{align}
  \epsilon & < 1.7 \times 10^{-5}
  \mathrm{\ for \ }
  T = 1 \; \mathrm{year \ and \ }
  N_s = 100
  ,
  \nonumber
  \\
  \epsilon & < 1.1 \times 10^{-5}
  \mathrm{\ for \ }
  T = 2 \; \mathrm{months \ and \ }
  N_s = 1000
  .
  \nonumber
\end{align}
So, we can relax the requirement on the relative precision of the
detector for an observation of $1$ second on a single source to
$10^{-5}$.
This is an acceptable value for nowaday instruments, considering that
the cosmic microwave background anisotropy of this order has already
been successfully mapped.

In the above formulation, a continuous observation is used.
This is, however, not necessary for such a measurement.
Consider the case there is a time interval between the measurement of
$S_1$ and $S_2$, for example, we first observe the target for a time
of $T_o$ and measure the fluence $S_1$, after a time interval $T_i$,
we observe it again for a time of $T_o$ and measure the fluence $S_2$.
It is easy to deduce
\begin{equation}
  \label{eq:Delta_lnd_measure_3}
  \left.
    \frac{\Delta d_L}{d_L}
  \right\vert_{\Delta t_0 = T_o + T_i}
  =
  \frac
  {
    S_1 - S_2
  }
  {
    S_1 + S_2
  }
  .
\end{equation}
This procedure can be repeated many times, such that we can sum up all
the $S_1$ and $S_2$, respectively, to increase the observation time.
As an example, let $T_i$ to be one year and $T_o$ to be a few hours.
We can observe the target at the same specific time of every day for
two years, then sum up the fluences of the first year and the second
year, respectively, as $S_1$ and $S_2$.
Thus, we can calculate $\Delta d_L / d_L$.

Anyway, if we do manage to successfully conduct this kind of
measurement, benefiting from the abundance of galaxies in the
universe, this method may be quite promising for measuring the
expansion rate of the universe.
Hopefully, we could even map the three-dimensional cosmic expansion,
i.e., the cosmic expansion rate of the different directions in the sky
along the redshift.
Also note that $\Delta d_L / d_L$ has the same order from the redshift
of zero to redshifts $z > 4$ (see Fig.~\ref{fig:dlnd}).
This covers the redshift range in which the dark energy plays its role
in the cosmic expansion, so the method may track the whole dynamics of
the dark energy.
In addition, the method not only directly measures the cosmic
expansion rate, but also is independent of any specific astrophysical
process, which is usually very complex and affected by many factors.
It is based on clean and clear fundamental physics, which strengthens
its robustness.

Compared to the Sandage-Loeb test, though the derivation of our method
is very similar to that of Sandage-Loeb test, the requirement on the
instrument, the observational target, and the expected outcome are
quite different.
The Sandage-Loeb test requires precise measurement of the redshift,
while our method requires precise measurement of the fluence.
The targets for the Sandage-Loeb test are quasars, while in our
method, the targets are galaxies, which are much more abundant in our
universe, thus could give more detailed information about the cosmic
expansion.
The redshift coverages of the two methods are also quite different.
As shown in~\cite{Corasaniti:2007bg}, the redshift coverage of the
Sandage-Loeb test is roughly between $2$ and $5$,
while our method covers the range from
a redshift close to zero (see discussions in the next section)
to redshifts $z > 4$.
All these make our method a novel one from the Sandage-Loeb test
despite the similar derivation.

In the above analyses, we have only considered the ideal condition,
for example, we have assumed the source rest in the comoving reference
frame and the photons propagate freely from source to the observer.
This is of course not the case of that in the real universe.
For the very high precision required by the method, impacts from the
peculiar velocity of the source and from the gravitational lensing
should be investigated seriously.
In the next section, we will show that the impacts from the peculiar
velocity can be safely ignored in a wide redshift range.
For the gravitational lensing, since we have so many galaxies in our
universe, we can expect that its impacts could, at least, be
eliminated by an average over the galaxies at the same redshift.
We leave it to future studies whether the gravitational lensing
will affect single measurements in our method.

\section{Peculiar velocity}

In this section, we study the impacts from the peculiar velocity of
the source on our method.

First, consider a isotropic point source and an observer nearby rest
in a local inertial reference frame. Assume the source has a velocity
of $\boldsymbol{v}$ with respect to the observer. Imagine a spherical
coordinate with the source as the origin and the direction of
$\boldsymbol{v}$ as the zenith direction. Say, the observer has an
inclination angle $\alpha$ in the coordinate. The momenta of the
photons received by the observer would have the same inclination angle
$\alpha$. But, to the source, as a result of the aberration of light
caused by the motion of the source, the momenta have a different
inclination angle $\bar{\alpha}$, which relates $\alpha$ through
\begin{equation}
  \label{eq:inclination_relation}
  \tan \frac{\alpha}{2}
  =
  \sqrt
  {
    \frac{1 - \beta}{1 + \beta}
  }
  \tan \frac{\bar{\alpha}}{2}
  ,
\end{equation}
where $\beta = v/c$.
So, the corresponding solid angles around above mentioned momenta have
different magnitudes to the observer and to the source. The ratio
between them is
\begin{equation}
  \label{eq:solid_angle_relation}
  \frac
  {
    \mathrm{d} \Omega
  }
  {
    \mathrm{d} \bar{\Omega}
  }
  =
  \frac
  {
    \sin \alpha
  }
  {
    \sin \bar{\alpha}
  }
  \frac
  {
    \mathrm{d} \alpha
  }
  {
    \mathrm{d} \bar{\alpha}
  }
  \equiv
  J(\alpha, \beta)
  .
\end{equation}
In addition to this, there is also the Doppler effect
\begin{equation}
  \label{eq:doppler_redshift}
  1 + z_d = (1 - \beta \cos \alpha) \gamma
  ,
\end{equation}
where $z_d$ is the Doppler redshift
and $\gamma = 1 / \sqrt{1 - \beta^2}$.
Taking into account both the aberration of light and the Doppler
effect, compared to the case of a rest source, the flux observed by
the observer is increased by a factor of
\begin{equation}
  \label{eq:flux_increase_factor}
  \frac{1}{(1 + z_d)^2}
  \cdot
  \frac{1}{J(\alpha, \beta)}
  =
  \frac
  {
    1
  }
  {
    [(1 - \beta \cos \alpha) \gamma]^2
    J(\alpha, \beta)
  }
  .
\end{equation}

Next, consider the cosmological case of a isotropic point source at
comoving distance $r$, with a luminosity of $L$, a peculiar velocity
of $\boldsymbol{v}$, and a cosmological redshift of $z_c$, and the
direction of $\boldsymbol{v}$ is at an angle of $\theta$ relative to
the line of sight (from the observer to the source).
The photons emitted by the source first experience an aberration and
the Doppler effect due to the velocity of the source, then the
cosmological redshift before they reach us.
As a result, the observed flux becomes
\begin{align}
  \label{eq:flux_p}
  F
  & =
  \frac{1}{4 \pi a_0^2 r^2}
  \frac{1}{(1 + z_c)^2}
  \frac{1}{(1 + z_d)^2}
  \frac{1}{J(\pi - \theta, \beta)}
  L
  \nonumber
  \\
  & =
  \frac{1}{4 \pi a_0^2 r^2}
  \frac{1}{(1 + z_c)^2}
  \frac{1}{[(1 + \beta \cos \theta) \gamma]^2}
  \frac{1}{J(\pi - \theta, \beta)}
  L
  ,
\end{align}
where $a_0 = a(t_0)$.
So, when taking into account the peculiar velocity of the source, the
corrected luminosity distance is given by
\begin{align}
  \label{eq:luminosity_distance_p}
  d_{L, p}
  & =
  a_0 r (1 + z_c)
  (1 + z_d)
  \sqrt
  {
    J(\pi - \theta, \beta)
  }
  \nonumber
  \\
  & =
  a_0 r (1 + z_c)
  (1 + \beta \cos \theta) \gamma
  \sqrt
  {
    J(\pi - \theta, \beta)
  }
  .
\end{align}

\begin{figure}[htbp]
  \centering
  \includegraphics[width = 0.5 \textwidth]{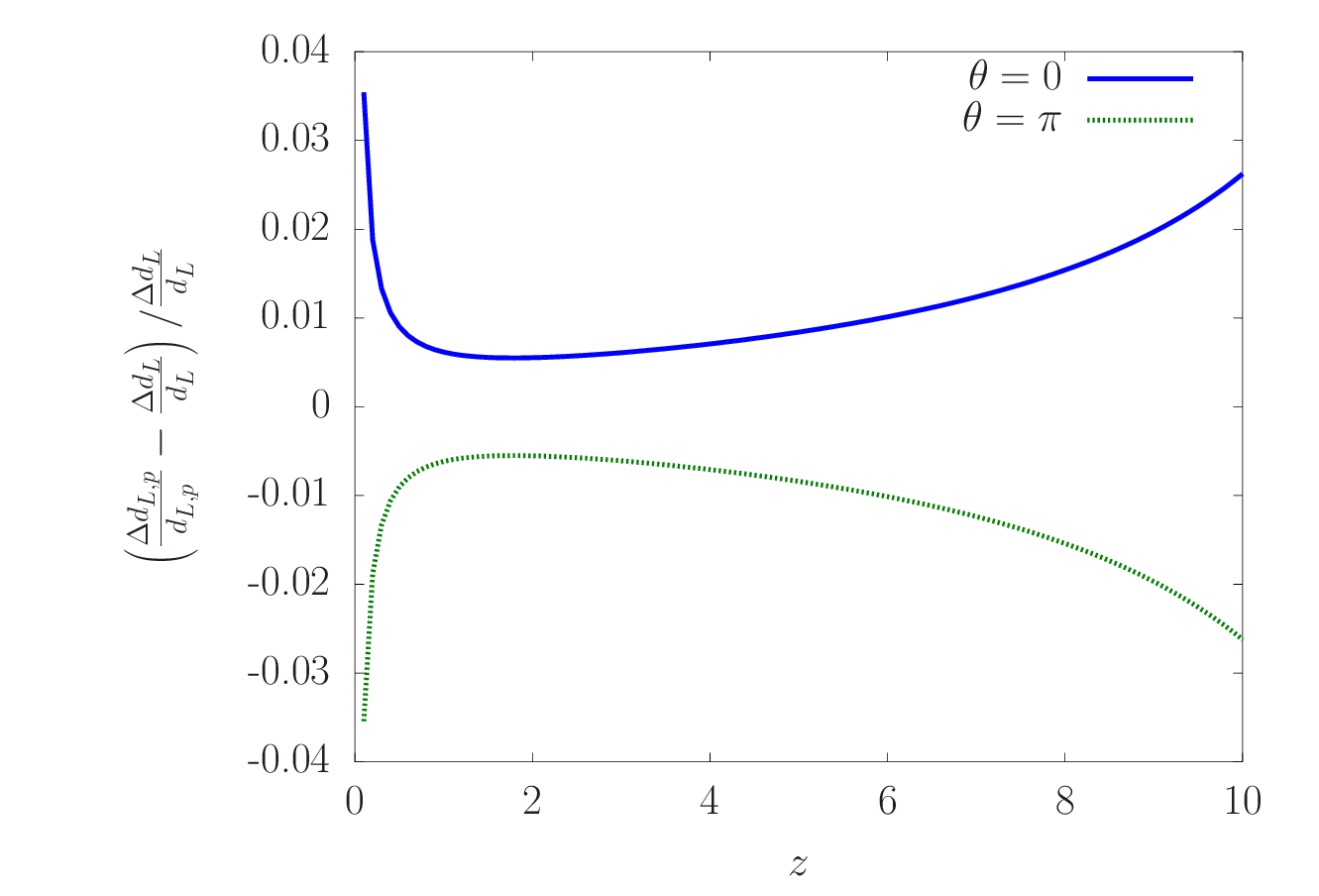}
  \caption
  {
    $\left( \frac{\Delta d_{L, p}}{d_{L, p}} - \frac{\Delta d_L}{d_L}
    \right) / \frac{\Delta d_L}{d_L}$
    versus redshift $z$
    for a peculiar velocity of
    $v = 1000 \ \mathrm{km} \, \mathrm{s}^{-1}$
    and the flat $\Lambda$CDM cosmological model
    with $\Omega_{m, 0} = 0.27$ and $H_0 = 70.5 \ \mathrm{km} \,
    \mathrm{s}^{-1} \, \mathrm{Mpc}^{-1}$.
  }
  \label{fig:relative_impact_of_v}
\end{figure}

It is easy to check that, under the condition of $\beta \ll 1$,
Eq.~(\ref{eq:inclination_relation}) and
Eq.~(\ref{eq:solid_angle_relation})
reduce to
\begin{align}
  \bar{\alpha} & = \alpha + \beta \sin \alpha
  ,
  \\
  J(\alpha, \beta) & = \frac{1}{(1 + \beta \cos \alpha)^2}
  ,
\end{align}
and Eq.~(\ref{eq:luminosity_distance_p}) reduces to
\begin{align}
  \label{eq:luminosity_distance_p_small_beta}
  d_{L, p}
  & =
  a_0 r (1 + z_c)
  (1 + z_d)
  /
  (1 - \beta \cos \theta)
  \nonumber
  \\
  & =
  a_0 r (1 + z_c)
  (1 + \beta \cos \theta) \gamma
  /
  (1 - \beta \cos \theta)
  .
\end{align}
Further ignore the change in the velocity of the source during the
observation, we have
\begin{equation}
  \label{eq:Delta_lnd_p}
  \frac{\Delta d_{L, p}}{d_{L, p}}
  =
  \frac{\Delta a_0}{a_0}
  + \frac{\Delta r}{r}
  + \frac{\Delta z_c}{1 + z_c}
  .
\end{equation}
Comparing the right hand side of this equation with that of
Eq.~(\ref{eq:Delta_lnd_1}), we can see that the first term is
unchanged.
The second term is newly introduced and is corresponding to the
distance the source moved through during the observation.
While the last term, at first glance, has the same form as the last
term of Eq.~(\ref{eq:Delta_lnd_1}), it actually includes two parts,
which may be called the time part and the space part.
The time part arises from that the redshift of a rest source changes
along the time due to the cosmic expansion, as was shown in
Eq.~(\ref{eq:Delta_z}).
The space part is in fact that the source has moved to a different
location during the observation due to its peculiar velocity and
sources at different comoving distances have different redshifts.
It is given by
\begin{equation}
  \label{eq:Delta_z_p}
  \left[
    \frac{\Delta z_c}{1 + z_c}
  \right]_p
  =
  \frac{H(z_c)}{c (1 + z_c)}
  \frac{a_0 \Delta r}{\sqrt{1 - k r^2}}
  ,
\end{equation}
where the subscript $p$ denotes the redshift drift caused by the
peculiar velocity, i.e., the space part.
So, the impact of the peculiar velocity on the change of the
luminosity distance is
\begin{align}
  \label{eq:impact_of_v}
  \frac{\Delta d_{L, p}}{d_{L, p}}
  -
  \frac{\Delta d_L}{d_L}
  & =
  \frac{\Delta r}{r}
  +
  \left[
    \frac{\Delta z_c}{1 + z_c}
  \right]_p
  \nonumber
  \\
  & =
  \left[
    1 +
    \frac{H(z_c)}{c (1 + z_c)^2}
    \frac{a_0 r (1 + z_c)}{\sqrt{1 - k r^2}}
  \right]
  \frac{\Delta r}{r}
  \nonumber
  \\
  & =
  \left[
    1 +
    \frac{H(z_c)}{c (1 + z_c)^2}
    \frac{d_L}{\sqrt{1 - k r^2}}
  \right]
  \frac{\Delta r}{r}
  ,
\end{align}
where
\begin{align}
  \label{eq:Delta_lnr}
  \frac{\Delta r}{r}
  & =
  \frac
  {
    (1 + z_c)^2 \sqrt{1 - k r^2}
  }
  {
    a_0 r (1 + z_c)
  }
  \frac
  {
    a(t_{\mathrm{em}}) \Delta r
  }
  {
    \sqrt{1 - k r^2}
  }
  \nonumber
  \\
  & =
  \frac
  {
    (1 + z_c)^2 \sqrt{1 - k r^2}
  }
  {
    d_L
  }
  v \cos \theta \Delta t_{\mathrm{em}}
  \nonumber
  \\
  & =
  \frac
  {
    (1 + z_c) \sqrt{1 - k r^2}
  }
  {
    d_L
  }
  v \cos \theta \Delta t_0
\end{align}
and
\begin{equation}
  \label{eq:curvature_term}
  \sqrt{1 - k r^2}
  =
  \sqrt
  {
    1 +
    \Omega_{k, 0}
    \frac{H_0^2}{c^2}
    \frac{d_L^2}{(1 + z_c)^2}
  }
  .
\end{equation}
We plot
$\left( \frac{\Delta d_{L, p}}{d_{L, p}} - \frac{\Delta d_L}{d_L}
\right) / \frac{\Delta d_L}{d_L}$
along the redshift in Fig.~\ref{fig:relative_impact_of_v}, from
which we can see that, from very low redshift to very high redshift,
the impact of the peculiar velocity is very small and can be safely
ignored.

\section{Summary}

In summary, basing on the definition of the luminosity distance and
its change along the time due to the cosmic expansion, we discussed
the possibility of directly measuring the expansion rate of the
universe through very precise measurement of the fluence.
Extremely stable sources are needed for the method.
We argued that galaxies may be chosen as the targets of the
observation.
Since composed of many stars, their luminosities are very stable.
Those involving violent astrophysical processes during the observation
could be easily identified and excluded.
Furthermore, the relative flux change caused by the cosmic expansion
has a flat-line spectrum, while those caused by other astrophysical
processes usually do not have such a character.
With this, we could further exclude those galaxies whose evolution
dominates over the cosmic expansion on the flux change.
We showed that, by simultaneously increasing the observation time and
physically adding the fluences from different galaxies,
the requirement on the relative precision of the detector for an
observation of $1$ second on a single galaxy can be relaxed to
$10^{-5}$.
We also showed that the peculiar velocity can be safely ignored for a
wide redshift range in our method.
The method not only directly measures the cosmic expansion rate, but
also is independent of any specific astrophysical process.
Benefiting from the abundance of galaxies in the universe,
the method may be quite promising.

\begin{acknowledgments}
  Shi Qi would like to thank the invitation of the 2009 KITPC program
  entitled ``Connecting Fundamental Physics with Observations''.
  In fact, the basic idea of this paper was originally triggered
  during the program.
  This research was supported by the National Natural Science
  Foundation of China under Grant No.~10973039,
  the Chinese Academy of Sciences under Grant No.~KJCX2-EW-W01,
  Jiangsu Planned Projects for Postdoctoral Research Funds 0901059C
  (for Shi Qi),
  the China Postdoctoral Science Foundation under Grant
  No.~20100471421 (for Shi Qi),
  and the Project of Knowledge Innovation Program (PKIP)
  of Chinese Academy of Sciences, Grant No.~KJCX2.YW.W10.
\end{acknowledgments}

\end{document}